\documentclass[a4paper,11pt]{article}
\usepackage{pos}

\usepackage{subcaption}

\usepackage{tikz}
\usetikzlibrary{positioning}
\usepackage{siunitx}
\usepackage{acro}
\acsetup{
  short-format={\scshape},
  foreign-format={\itshape},
}
\DeclareAcronym{crp}{
  short=CRP,
  long= Charge Readout Plane
}
\DeclareAcronym{cru}{
  short=CRU,
  long= Charge Readout Unit
}
\DeclareAcronym{surf}{
  short=SURF,
  long= Sanford Underground Research Facility
}
\DeclareAcronym{tpc}{
  short=TPC,
  long= Time Projection Chamber
}
\DeclareAcronym{pcb}{
  short=PCB,
  long= printed circuit board
}
\DeclareAcronym{cern}{
	short=cern,
	long= European Organization for Nuclear Research,
	foreign=\textit{Conseil européen pour la recherche nucléaire},
}
\DeclareAcronym{dune}{
  short=DUNE,
  long= Deep Underground Neutrino Experiment
}
\DeclareAcronym{lar}{
  short=\textup{LAr},
  long= liquid-Argon
}
\DeclareAcronym{fd}{
  short=FD,
  long= far detector
}
\DeclareAcronym{vd}{
  short=VD,
  long= Vertical Drift
}
\DeclareAcronym{hd}{
  short=HD,
  long= Horizontal Drift
}

\renewcommand{\logo}{\relax}

\title{The \acs{dune} vertical drift \acs{tpc}}
\manuallySeparateAuthors
\author*{Oliver Lantwin}
\author{for the DUNE collaboration}

\affiliation{Laboratoire d'Annecy de Physique des Particules,\\
  9 Chemin de Bellevue, Annecy, France}

\emailAdd{oliver.lantwin@lapp.in2p3.fr}

\abstract{
  The \acs{dune} experiment is a future long-baseline neutrino oscillation experiment aiming at measuring the neutrino CP-violating phase and establishing the neutrino mass hierarchy,
  as well as at a rich physics programme from supernovae over low-energy physics to beyond Standard Model searches.

  The baseline technology for the first far detector is a proven single-phase horizontal-drift liquid-Argon \acs{tpc} based on standard wire-chamber technology.

  For the second far detector,
  a new technology,
  the so-called ``vertical drift'' \acs{tpc} is currently being developed:
  It aims at combining the strengths of the two technologies tested in the \mbox{ProtoDUNE} cryostats at the \acs{cern} neutrino platform,
  the proven horizontal-drift single-phase TPC and the ambitious vertical-drift dual-phase TPC,
  into a single design,
  a vertical-drift single-phase liquid-Argon TPC using a novel perforated-PCB anode design.
  This design maintains excellent tracking and calorimetry performance while significantly simplifying the complexity of the TPC construction.

  This paper introduces the concept of the vertical drift \acs{tpc},
  presents first results from small-scale prototypes and a first full-scale anode module,
  and outlines the plans for future prototypes and the next steps towards the full second DUNE far detector.
}

\FullConference{%
  41st International Conference on High Energy physics - ICHEP2022\\
  6-13 July, 2022\\
  Bologna, Italy
}

\begin{document}
\maketitle

\section{Introduction}

After the discovery of neutrino oscillations,
an extensive experimental programme of solar,
reactor and accelerator neutrino experiments has measured many of the parameters of the neutrino-mixing matrix.
However, two important puzzle pieces remain:
While two squared-mass differences have been established,
two mass orderings of the neutrino mass states remain possible.
Additionally,
a CP-violating phase is allowed in the mixing matrix,
which could be part of the explanation of the baryon-asymmetry of the universe.
DUNE,
a next-generation long-baseline neutrino oscillation experiment,
aims to measure these parameters,
alongside a diverse physics programme ranging from supernova neutrinos and solar neutrinos, to beyond Standard Model measurements and nucleon decay studies\cite{DUNE:2020ypp}.
The experiment uses a neutrino beam from Fermilab, which is sent to far detectors at \ac{surf} \SI{1.5}{\kilo\metre} underground, for a baseline of \SI{1300}{\kilo\metre}\cite{DUNE:2020lwj}.
The DUNE collaboration consists of over 1300 scientists and engineers from 37 countries and \acs{cern}.

For its far detectors, DUNE uses \ac{lar} \acp{tpc}\cite{Rubbia:1977zz}.
\Ac{lar} provides a dense,
pure medium with prompt scintillation light for triggering using separate photo-detectors,
allowing the construction of \si{\kilo\tonne}-scale detectors,
while being much more abundant and affordable than Xenon.
\Ac{lar} \acp{tpc} offer fine-grained millimetric three-dimensional tracking and total-absorption calorimetry,
which allows identifying particles via energy loss and topology.

The baseline technology for the first \ac{dune} \acf{fd} module is a horizontal-drift single-phase \ac{lar} \ac{tpc} built using wire-chamber technology\cite{DUNE:2020txw},
as used by several previous experiments.
The single-phase \ac{vd} \ac{tpc} was chosen as the technology for the second far detector, \ac{fd}2.
With \SI{17.5}{\kilo\tonne} each,
the \ac{dune} \ac{fd} modules will be the largest \ac{lar} \acp{tpc} ever built.
A phased approach is foreseen,
with two far detectors for Phase I,
and two more \acp{fd} for Phase II,
for which the technology R\&D is ongoing.
Upgrades to the beam,
increasing the power from \num{1.2} to \SI{2.4}{\mega\watt},
and more capable near detectors are foreseen for the second phase.

\section{The \acl{vd} concept}

Since 2018 the two ProtoDUNE cryostats were used to test the \ac{dune} \ac{fd} technologies.
ProtoDUNE-SP successfully validated the \acf{hd} technology of \ac{fd}1,
while ProtoDUNE-DP tested the more ambitious dual-phase technology including signal amplification,
simpler construction and a longer drift-length.
The ProtoDUNE detectors demonstrated very good \ac{lar} purity,
allowing for a long \SI{6.5}{\metre} drift distance,
and resulting in excellent signal-to-noise ratio,
which meant that the gain from the gaseous phase of the dual phase technology was not needed.
However, some of the other advantages of ProtoDUNE-DP inspired the single-phase \ac{vd} technology,
which takes the best properties of both ProtoDUNE detectors for an improved \emph{single-phase} \ac{tpc}.

Based on the experience with the ProtoDUNE detectors at the \acs{cern} neutrino platform,
the new ``vertical drift'' technology was developed for the second \ac{dune} \ac{fd}.
It benefits from the advantages of the dual phase ProtoDUNE design
while eliminating the complexity introduced by the liquid-gas interface.

\begin{figure}
  \centering
  \includegraphics[width=0.5\textwidth]{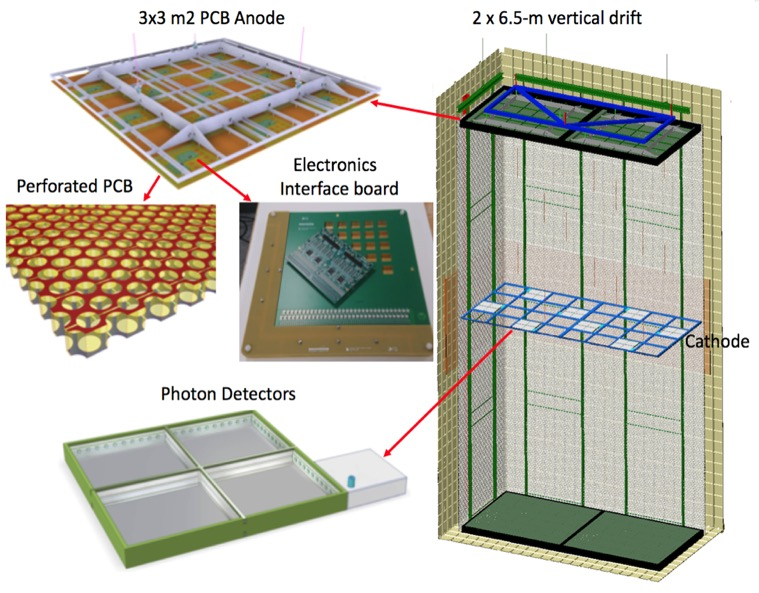}
  \caption{A cross-section of a single vertical drift module.}
  \label{fig:vd}
\end{figure}

A cross-section of a single vertical drift module is shown in Fig.~\ref{fig:vd}.
The cryostat design is shared with the first \ac{fd} module.
Unlike its wire-based anode however,
\ac{vd} uses anodes consisting of stacked segmented and perforated \acp{pcb}
with etched electrodes,
which are mechanically robust and modular for easy assembly,
and mass producible.
The cathode will be suspended at mid-height.
Finally,
photon detectors based on the X-ARAPUCA technology\cite{Machado:2018rfb} will be embedded in the cathode and cryostat walls for timing and triggering.

\begin{figure}
  \centering
  \includegraphics[width=0.6\textwidth]{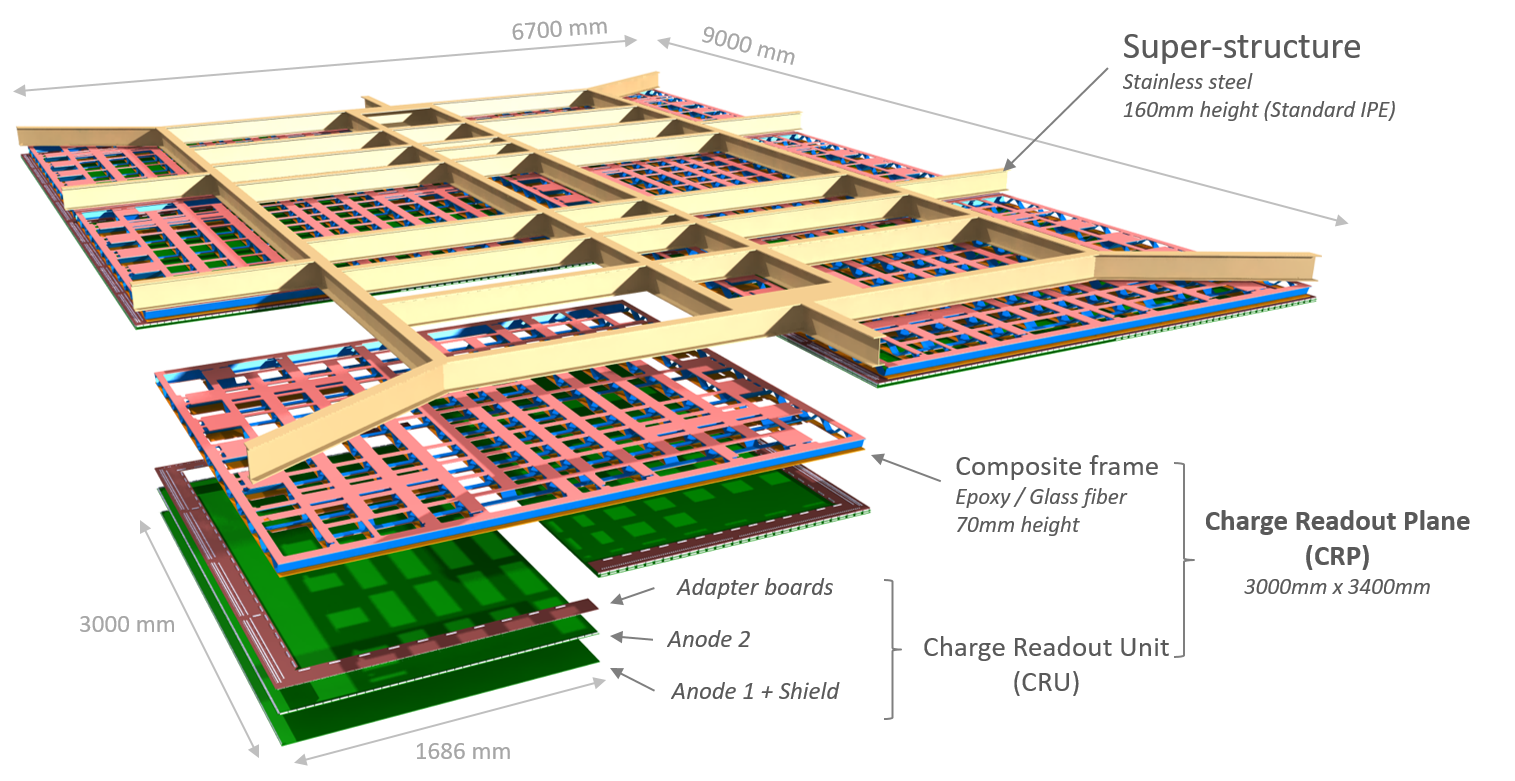}
  \caption{A schematic of a far detector \ac{crp}, attached to a mechanical support structure.}
  \label{fig:fd_crp}
\end{figure}

\begin{figure}
  \centering
  \subfloat[][]{
    \includegraphics[width=0.3\textwidth]{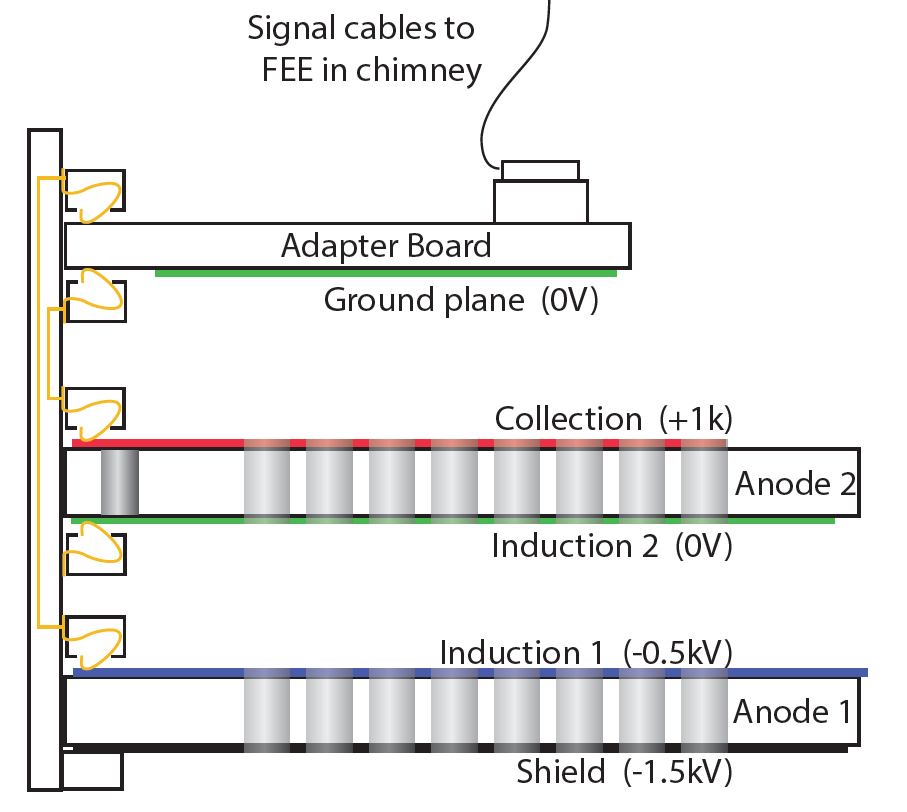}
    \label{fig:stack}
  }
  \subfloat[][]{
    \includegraphics[width=0.4\textwidth]{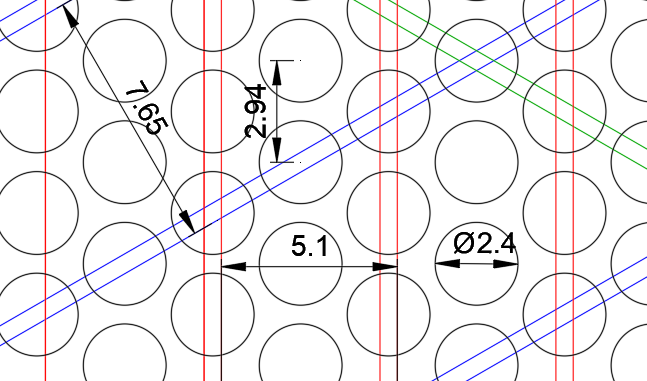}
    \label{fig:pattern}
  }
  \caption{(\subref{fig:stack}) A side-on view of the \ac{pcb} stack. (\subref{fig:pattern}) The strip pattern on the anode planes.}
  \label{fig:stackpattern}
\end{figure}

A single \ac{crp} is shown in Fig.~\ref{fig:fd_crp} in an exploded view.
Up to six \acp{crp} are going to be attached to the same mechanical support structure,
which is suspended from the cryostat roof by four cables,
in the case of the top \acp{crp}.
Each \ac{crp} consists of two \acp{cru},
each consisting of two anode planes.
The two \acp{cru} are attached to a single composite frame with the same thermal expansion coefficient as the PCBs,
and which is mechanically decoupled from the mechanical support frame to allow independent thermal contraction and expansion.
These anode planes are connected via edge connectors to the adapter boards as shown in Fig.~\ref{fig:stack}.
The individual strips are etched into the anode planes following the pattern shown in Fig.~\ref{fig:pattern}.

Two different readout systems are used for the \acp{crp} at the top and bottom of the cryostat. 
The top readout electronics are fully accessible from the top,
which allows for simple access for maintenance and upgrade of the electronics while the detector is filled.
They are connected to the adapter boards via KEL connectors.
The bottom readout electronics however are near the cryostat floor,
mounted directly onto the adapter boards on the underside of the \acp{crp},
mounted directly onto the adapter boards,
and are fully immersed in \ac{lar}.
While the top electronics are based on the ProtoDUNE-DP electronics,
the bottom electronics design is shared with \ac{fd}-\acs{hd}.

\begin{figure}
  \centering
  \includegraphics[width=0.8\linewidth]{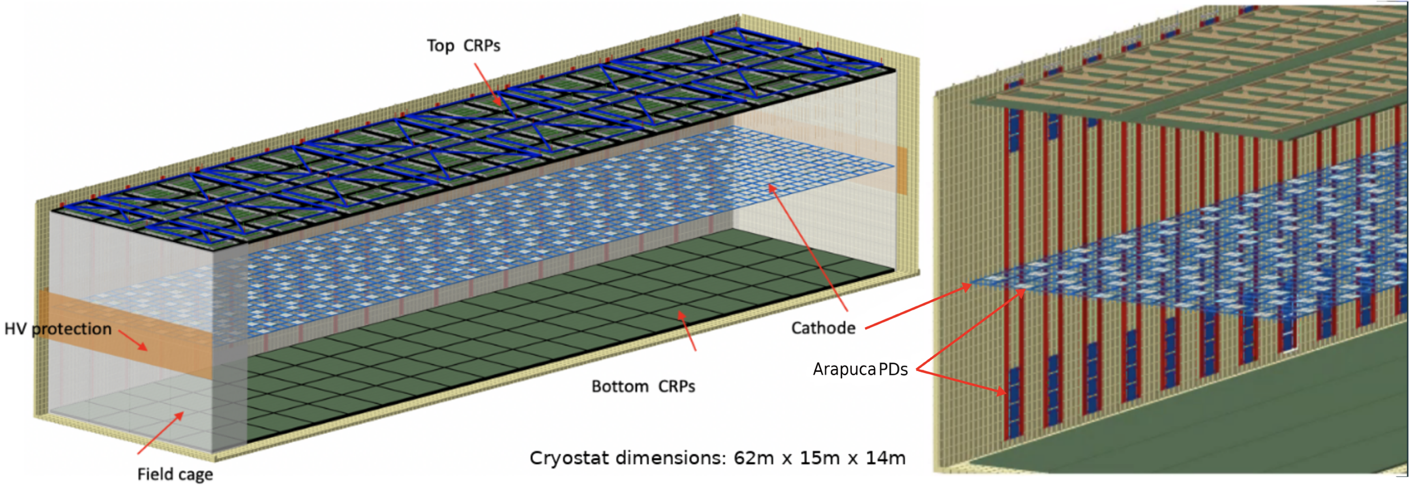}
  \caption{The layout of the Vertical Drift \ac{fd} module.}
  \label{fig:fdvd}
\end{figure}

The final \SI{17.5}{\kilo\tonne} \ac{fd}2 \ac{vd} will have 80 \acp{crp} at the top and 80 \acp{crp} at the bottom of the cryostat,
each measuring $\SI{3.4}{\metre}\times \SI{3}{\metre}$,
as shown in Fig.~\ref{fig:fdvd}.
The \ac{fd}-component mass production should start in 2024.

\section{Prototyping the \ac{vd} \ac{tpc}}

Since the conception of the \ac{vd} technology,
several milestones needed to be reached on the way to the full \ac{fd}.
After initial tests using a 50 litre proof-of-concept \ac{tpc},
a full-\ac{crp} test in a cold-box demonstrated the concept,
leading to the full Module-$0$,
currently in preparation,
to demonstrate the technology's readiness for the far detector.

\subsection{The 50 litre \ac{tpc}}

A 32$\times\SI{32}{\centi\metre\squared}$ prototype \ac{tpc} was built at \acs{cern} to test hole-sizes,
strip pitch,
signal shapes and energy resolution using cosmic muons and a ${}^{207}$Bi source in several runs from \numrange{2020}{2022}.

Several different \ac{pcb} configurations were tested,
including a single \ac{pcb} with two views,
and two stacked \acp{pcb} with two induction and one readout view for three views in total,
in addition to a shield layer.
The prototype was also used to successfully test the edge connectors used for the Module-$0$ \acp{crp}.
For all these tests,
the \SI{50}{\litre} \ac{tpc} was read out using bottom readout electronics.

\subsection{The cold-box}

\begin{figure}
  \centering
  \subfloat[][]{
    \includegraphics[width=0.66\textwidth]{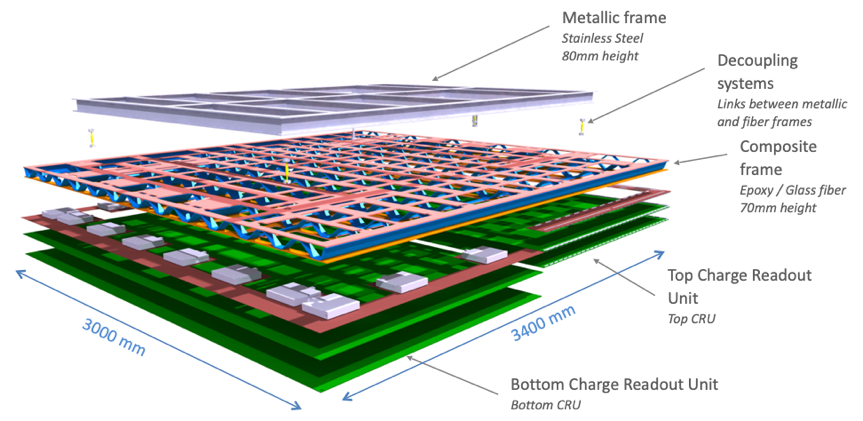}
    \label{fig:crp}
  }
  \subfloat[][]{
    \includegraphics[width=0.33\textwidth]{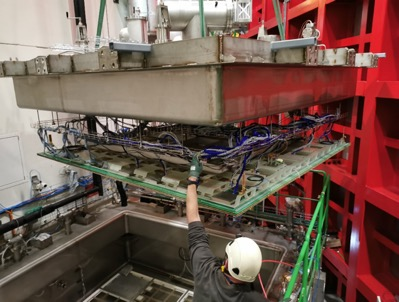}
    \label{fig:coldbox}
  }
  \caption{(\subref{fig:crp}) A diagram of the first full \ac{crp}, instrumented with different readout systems for each half. (\subref{fig:coldbox}) The first \ac{crp} as it is being lowered into the cold-box.}
\end{figure}

The $4\times4\times\SI{1}{\metre\cubed}$ NP02 cold-box at the \acs{cern} neutrino platform was refurbished in order to test full-scale \ac{crp} modules,
the cathode and the photon detection system at cold with a drift distance of about \SI{20}{\centi\metre}.
Half of the first \ac{crp},
built in 2021 and shown in Fig.~\ref{fig:crp},
is instrumented half with top,
half with bottom electronics to test both readout electronic systems.

Due to manufacturing constraints,
each anode panel has to be assembled from 6 segments to form a $\SI{3.4}{\metre}\times \SI{1.5}{\metre}$ panel,
which are glued together with epoxy in a half-lap configuration.
Channels are bridged between segments using screen-printed silver-epoxy connections,
which are then cured using heating strips.

\begin{figure}
  \centering
  \subfloat[][]{
    \includegraphics[width=0.6\textwidth]{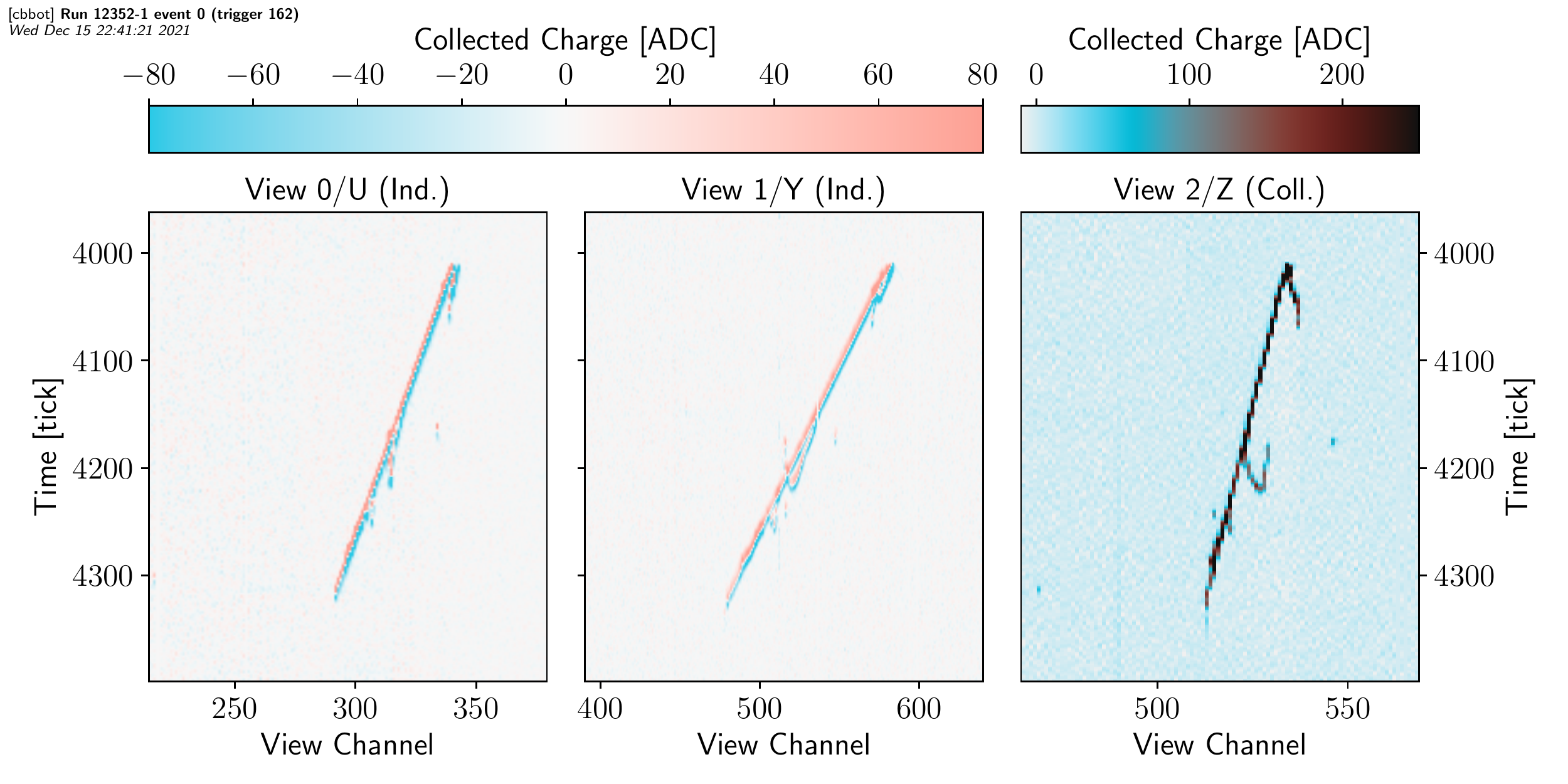}
    \label{fig:event}
  }\\
  \subfloat[][]{
    \includegraphics[width=0.8\textwidth]{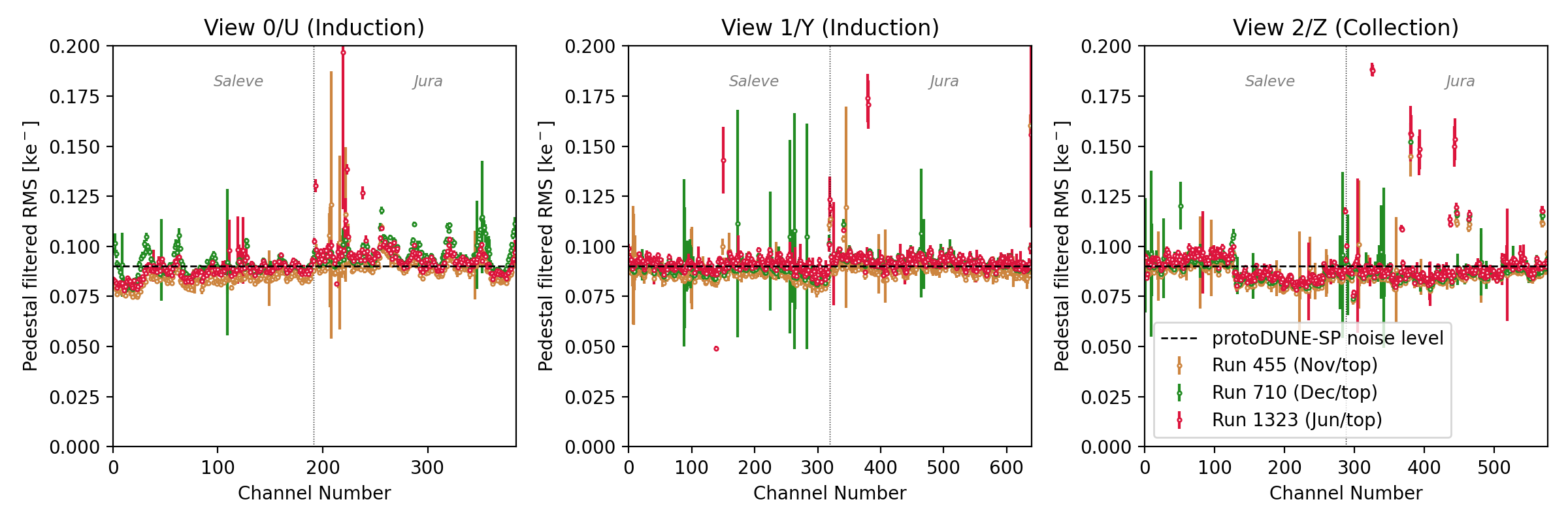}
    \label{fig:noise}
  }
  \caption{(\subref{fig:event}) A reconstructed track from a cosmic ray event in the cold-box.
    (\subref{fig:noise}) The noise levels achieved,
    compared to the reference noise level of the \acl{hd} TPC.}
\end{figure}

The \ac{crp} design was successfully validated at cold and both the gluing and the interconnection of segments was demonstrated.
Fig.~\ref{fig:coldbox} shows the \ac{crp} being lowered into the cold-box.
Two runs with large samples of $\mathcal{O}(10^6)$ triggers each were taken in November and December 2021,
with full analysis in progress.
Good tracks were seen in both readout systems,
as shown in Fig.~\ref{fig:event}.
An excellent signal-to-noise ratio was also demonstrated,
matching that of the \ac{hd} technology,
as shown in Fig.~\ref{fig:noise}.

\subsection{ProtoDUNE-VD}

The NP02 ProtoDUNE cryostat will be re-instrumented as Module-$0$ of the \ac{fd}-\ac{vd} for 2023,
with dedicated test beams and cosmic runs in 2023 and 2024.
It will contain two full modules with two top and bottom \ac{crp} each,
and suspended cathodes in between,
resulting in a drift distance of about \SI{3}{\metre},
due to the limited size of the cryostat.
Several more cold-box runs are being performed this year to test the final strip orientation  of \SI{\pm30}{\degree} and \SI{90}{\degree},
the edge connectors and homogeneous top/bottom modules,
and to test the \acp{crp} before integration into the Module-$0$.

\section{Conclusion}

The \acl{vd} technology aims to unite the best features of both ProtoDUNE technologies for the second \ac{dune} \acl{fd}.
It maintains the high performance and signal-to-noise ratio of the \acl{hd} technology,
while being more mechanically robust and simple to assemble.
The prototyping is progressing well and the first parts of the Module-$0$ are assembled and being tested.
The completion of Module-$0$ is foreseen for early 2023, on track for \ac{dune} Phase I.

\bibliographystyle{JHEP}
\bibliography{bibliography}

\end{document}